\documentclass[12pt]{article}
\usepackage{geometry}                
\usepackage{graphicx}
\usepackage{bm}
\usepackage[dvips]{color}
\usepackage{amssymb,amsmath,amsthm,amssymb}
\usepackage{mathrsfs} 
\usepackage{epstopdf}
\def\TODAY{4 January 2008}

\title{\bf Bounding the Bogoliubov coefficients}
\author{{\Large Petarpa Boonserm and Matt Visser}\\[10pt]
School of Mathematics, Statistics, and Computer Science\\
Victoria University of Wellington\\
New Zealand}
\date{\TODAY; \LaTeX-ed \today}                                           

\begin{document}
\maketitle
\def\d{{\mathrm{d}}}
\newcommand{\scri}{\mathscr{I}}
\newcommand{\sun}{\ensuremath{\odot}}
\def\tr{{\mathrm{tr}}}
\def\sech{{\mathrm{sech}}}
\def\etc{\emph{etc}}
\newcommand{\pacomment}[1]{ {\em \color{blue} #1}}

\def\lint{\hbox{\Large $\displaystyle\int$}} 
\def\hint{\hbox{\Huge $\displaystyle\int$}}  

\begin{abstract}
While over the last century or more considerable effort has been put into the problem of finding \emph{approximate solutions} for wave equations in general, and quantum mechanical problems in particular, it appears that as yet relatively little work seems to have been put into the complementary problem of establishing rigourous \emph{bounds} on the exact solutions. We have in mind either bounds on parametric amplification and the related quantum phenomenon of particle production (as encoded in the Bogoliubov coefficients), or bounds on transmission and reflection coefficients. Modifying and streamlining an approach developed by one of the present authors [Phys.\ Rev.\  A {\bf59}  (1999) 427--438], we investigate this question by developing a \emph{formal} but \emph{exact} solution for the appropriate second-order linear ODE in terms of a  time-ordered exponential of $2\times2$ matrices, then relating the Bogoliubov coefficients to certain invariants of this matrix. By bounding the matrix in an appropriate manner, we can thereby bound the Bogoliubov coefficients.

\null \qquad\qquad\quad Reflection coefficient,  rigorous bounds.\\
arXiv:0801.0610v1 [quant-ph]
\end{abstract}

\clearpage
\section{Introduction}

There are numerous  physical situations in which it is both extremely interesting and important to study the second-order ODE~\cite{bounds}
\begin{equation}
\ddot \phi(t) + \omega^2(t) \,\phi(t)=0,
\label{E:time}
\end{equation}
or its equivalent in the space domain~\cite{bounds}
\begin{equation}
 \phi''(x) + k^2(x) \,\phi(x)=0.
 \label{E:space}
\end{equation}
Viewed in terms of the time domain, equation (\ref{E:time}) can be viewed as an example of parametrically excited oscillation; it arises for instance when a wave propagates through a medium whose refractive index is externally controlled to be a function of time (though remaining spatially invariant).\footnote{For instance, situations of this type have been used to model sonoluminescence~\cite{sonoluminescence}, and more recently both quasiparticle production in analogue spacetimes~\cite{quasiparticle} and analogue signature change events~\cite{signature}.  In all these situations it is extremely useful to have rigorous and largely model-independent bounds on the amount of particle production that might reasonably be expected.}\; In contrast, the spatial version of this equation as presented in (\ref{E:space}) arises classically in situations where the refractive index is spatially dependent (so called ``index gradient'' situations), or in a quantum physics context when considering the Schrodinger equation for a time-independent potential:
\begin{equation}
- {\hbar^2\over2m} \, \phi''(x) + V(x) \,\phi(x)=E \,\phi(x),
\end{equation}
as long as one makes the translation
\begin{equation}
k^2(x) \leftrightarrow {2m[E-V(x)]\over \hbar^2}.
\end{equation}
However they arise, equations (\ref{E:time}) and (\ref{E:space}) are central to the study of both quantum physics and wave phenomena generally.

Because of this central importance, over the last century or more a vast body of work has gone into the question of finding \emph{approximate solutions} to equations  (\ref{E:time}) and (\ref{E:space}), most typically based on JWKB techniques and their variants (phase integral techniques, \etc.)~\cite{approximate}.  In contrast very little work seems to have gone into the physically important question of finding \emph{explicit bounds} on the relevant Bogoliubov coefficients and/or reflection and transmission coefficients~\cite{bounds}. 

In the current article we shall modify and streamline the analysis of~\cite{bounds}; presenting an alternative proof that is considerably more direct and focussed than that in~\cite{bounds}.
To keep the discussion simple and straightforward we shall assume that $\omega(t)\to\omega_0$ (equivalently $k(x)\to k_0$) outside some region of compact support $[t_i,t_f]$ (equivalently $[x_i,x_f]$). That is, concentrating on the time-domain formulation of equation (\ref{E:time}), the quantity $\omega^2(t)-\omega_0^2$ is a function of compact support.\footnote{This ``compact support'' condition is not  strictly necessary, and at the cost of a little more analysis one can straightforwardly extend the comments below to a situation where there is a finite limit $\omega(t)\to\omega_{\infty}$ as $t\to\pm\infty$~\cite{bounds}. At the cost of somewhat more tedious additional work,  there are also useful things that can be said of the situation where $\omega(t)\to\omega_{\pm\infty}$, with $\omega_{-\infty}\neq \omega_{+\infty}$, as $t\to\pm\infty$~\cite{bounds}.} Because of this compact support property we know that everywhere outside the region $[t_i,t_f]$ the exact solution of the wave equation (\ref{E:time}) is given by linear combinations of $\exp(\pm i \omega_0\, t)$, and that the central question to be investigated is the manner in which exact solutions on the initial domain $(-\infty,t_i)$ ``connect'' with exact solutions on the final domain $(t_f,+\infty)$. Describing and characterizing this ``connection'' is exactly what the Bogoliubov coefficients are designed to do.

\section{Time-ordered exponentials}
\def\txt{\textstyle}
\def\lint{\hbox{\Large $\displaystyle\int$}} 
\def\hint{\hbox{\Huge $\displaystyle\int$}} 
We are interested in solving, \emph{exactly} but possibly \emph{formally}, the second-order PDE
\begin{equation}
\ddot \phi(t) + \omega^2(t) \,\phi(t)=0.
\end{equation}
One way of proceeding is as follows: Define a momentum
\begin{equation}
\pi  = \dot\phi,
\end{equation}
and then rewrite the second-order ODE as a system of first-order ODEs
\begin{equation}
\dot \phi = \pi;
\end{equation}
\begin{equation}
\dot\pi = -\omega^2(t)\; \phi;
\end{equation}
or in matrix notation (where we have carefully arranged all matrix elements and vector components to carry the same engineering dimensions)  
\begin{equation}
{\d\over\d t} \left[\begin{array}{c} \phi\\ \pi/\omega_0 \end{array}\right] =
\left[\begin{array}{cc} 0 & \omega_0 \\ -\omega^2/\omega_0 & 0 \end{array}\right] \; 
\left[\begin{array}{c} \phi\\ \pi/\omega_0 \end{array}\right].
\end{equation}
This matrix ODE \emph{always} has a \emph{formal} solution in terms of the so-called ``time ordered exponential'' 
\begin{equation}
\left[\begin{array}{c} \phi\\ \pi/\omega_0 \end{array}\right]_t = 
\mathscr{T} \left\{ \exp\left( 
\int_{t_0}^t 
\left[\begin{array}{cc} 0 & \omega_0\\ -\omega^2(\bar t) /\omega_0& 0 \end{array}\right] 
\d \bar t
\right) \right\} \; \left[\begin{array}{c} \phi\\ \pi/\omega_0 \end{array}\right]_{t_0}.
\end{equation}
The meaning of the time-ordered exponential is somewhat tricky, but ultimately is just a $2\times2$ matrix specialization of the operator-valued version of the ``time ordered exponential''  familiar from developing quantum field theoretic perturbation theory in the so-called ``interaction picture''~\cite{qft-interaction}. Specifically,  let us partition the interval $(t_0,t)$ as follows:
\begin{equation}
t_0 < t_1 < t_2 < t_3 ... < t_{n-3} < t_{n-2} < t_{n-1} < t_n=t,
\end{equation}
and define the ``mesh'' as 
\begin{equation}
M = \max_{i\in (1,n)} \{ t_i - t_{i-1} \}.
\end{equation}
Then define the time-ordered exponential as
\begin{eqnarray}
T(t) &=& \mathscr{T} \left\{ \exp\left( 
\int_{t_0}^t 
\left[\begin{array}{cc} 0 & \omega_0\\ -\omega^2(\bar t)/\omega_0 & 0 \end{array}\right] 
\d \bar t
\right) \right\} 
\nonumber\\
&\equiv&
\lim_{M\to 0,\; (n\to \infty)} \prod_{i=0}^{n-1}  
\exp\left( 
\left[\begin{array}{cc} 0 & \omega_0\\ -\omega^2(t_{n-i})/\omega_0 & 0 \end{array}\right]  
\; (t_{n-i}-t_{n-i-1}) 
\right).
\end{eqnarray}
Note that in this matrix product ``late times'' are always ordered to the left, and ``early times'' to the right.
By working with this time-ordered matrix  we will be able to extract \emph{all} the interesting physics. (If we work in the space domain then the equivalent matrix $T$ is ``path-ordered'', and is closely related to the so-called  ``transfer matrix''.)

\begin{itemize}
\item
Since all of the  ``complicated'' physics takes place for $t\in (t_i,t_f)$, it is also useful to define
\begin{equation}
T = \mathscr{T} \left\{ \exp\left( 
\int_{t_i}^{t_f} \left[\begin{array}{cc} 0 & \omega_0 \\ -\omega^2(\bar t)/\omega_0 & 0 \end{array}\right] \d \bar t
\right) \right\} =
 \left[\begin{array}{cc} a & b\\ c & d \end{array}\right].
 \end{equation}

\item 
We are guaranteed that $\det[T]=1$, that is $ad-bc=1$. This follows from the fact that $\det[T] = \exp\{\tr(\ln[T])\}$, and the explicit formula for $T$ above.

\item
 Another particularly nice feature is that with the current definitions the transfer matrix $T$ is manifestly \emph{real}. This is relatively rare when setting up scattering or particle production problems, so we shall make the most of it.
 \end{itemize}
 
\section{Bogoliubov coefficients}

Let is now calculate the Bogoliubov coefficients.  Before $t_i$, and after $t_f$, the wave-function is just linear combinations of $\exp(\pm i \omega_0 \, t)$. We can prepare things so that before $t_i$ the wavefunction is pure  $\exp(+ i \omega_0 \, t)$,
\begin{equation}
\psi(t\leq t_i) = \exp(+ i \omega_0 \, t);
\end{equation}
in which case after $t_f$ the wavefunction will be a linear combination
\begin{equation}
\psi(t\geq t_f) = \alpha \exp(+ i \omega_0 \, t) + \beta  \exp(- i \omega_0\, t) ,
\end{equation}
where the Bogoliubov coefficients $\alpha$ and $\beta$ are to be calculated. 
That is,  we have
\begin{equation}
\left[\begin{array}{c} \phi\\ \pi/\omega_0 \end{array}\right]_{t_i} =
\left[\begin{array}{c}  \exp(+ i \omega_0 \, t_i) \\  i  \exp(+ i \omega_0 \, t_i) \end{array}\right],
\end{equation}
and
\begin{equation}
\left[\begin{array}{c} \phi\\ \pi/\omega_0 \end{array}\right]_{t_f} =
\left[\begin{array}{c}  \alpha \exp(+ i \omega_0 \,t_f) + \beta  \exp(- i \omega_0 \, t_f) \\  
i \left\{ \alpha \exp(+ i \omega_0 \, t_f) - \beta  \exp(- i \omega_0 \, t_f)  \right\}\end{array}\right].
\end{equation}
But we also have
\begin{equation}
\left[\begin{array}{c} \phi\\ \pi/\omega_0 \end{array}\right]_{t_f} 
= T \left[\begin{array}{c} \phi\\ \pi/\omega_0 \end{array}\right]_{t_i},
\end{equation}
implying 
\begin{equation}
\left[\begin{array}{c}  \alpha \exp(+ i \omega_0\, t_f) + \beta  \exp(- i \omega_0 \,t_f) \\  
i  \left\{ \alpha \exp(+ i \omega_0 \,t_f) - \beta  \exp(- i \omega_0 \,t_f)  \right\}\end{array}\right]
=
\left[\begin{array}{c}  a \exp(+ i \omega_0 \, t_i) + b\, i   \exp(+ i \omega_0 \, t_i)\\  
c  \exp(+ i \omega_0 \, t_i) +d\, i  \exp(+ i \omega_0 \, t_i) \end{array}\right].
\end{equation}
Solving these simultaneous linear equations we find
\begin{equation}
\alpha = {1\over2} \left[ a+d + i\left(b-{c}\right)\right] \;\exp(-i \omega_0\,[t_f - t_i]),
\end{equation}
\begin{equation}
\beta = {1\over2} \left[ a-d + i\left(b+{c}\right)\right] \;\exp(-i \omega_0\,[t_f + t_i]),
\end{equation}
so that the Bogoliubov coefficients are simple linear combinations of elements of the matrix $T$.
Then (remember the matrix $T$ is \emph{real})
\begin{equation}
|\alpha|^2 = {1\over4} \left\{ 
 (a+d)^2 + (b-c)^2 
\right\},
\end{equation}
\begin{equation}
|\beta|^2 = {1\over4} \left\{ 
 (a-d)^2 + (b+c)^2 
\right\},
\end{equation}
and so
\begin{eqnarray}
|\alpha|^2-|\beta|^2 &=& { (a+d)^2 + (b-c)^2 - (a-d)^2  -(b+c)^2   \over4},
\\
&=&{2ad -2bc +2ad - 2bc\over 4} = {ad-bc} = 1,
\end{eqnarray}
thus verifying  that, (thanks to the unit determinant condition), the Bogoliubov coefficients are properly normalized.  Particle production is governed by the $\beta$ coefficient in the combination
\begin{eqnarray}
|\beta|^2 &=& {1\over 4} \left\{ (a-d)^2 + \left(b+{c}\right)^2 \right\},
\\
&=&  {1\over 4} \left\{ a^2 + d^2 -2 ad + b^2  + {c^2} + 2bc   \right\},
\\
&=&  {1\over 4} \left\{ a^2 + d^2+ b^2  + {c^2} - 2 \right\},
\\
&=& {1\over4} \tr\{ T \, T^T - I\}.
\end{eqnarray}
Note that the transpose $T^T$ is now \emph{time-anti-ordered}.

Similarly
\begin{eqnarray}
|\alpha|^2 &=& {1\over4} \left\{  (a+d)^2 + (b-{c})^2  \right\},
\\
&=& {1 \over 4} \left\{a^2 + d^2 +2 ad + b^2 + {c^2} - 2bc\right\},
\\
&=&  {1\over 4} \left\{ a^2 + d^2+ b^2 + {c^2} + 2 \right\},
\\
&=& {1\over4} \; \tr\{ T \, T^T + I\}.
\end{eqnarray}
In summary, we can always formally solve the relevant ODE, either equation (\ref{E:time}) or its equivalent equation (\ref{E:space}), in terms of the time-ordered exponential, and we can always formally extract the Bogoliubov coefficients in terms of traces of the form $\tr\{T\,T^T\}$. We shall now use these formal results to derive rigorous bounds on the Bogoliubov coefficients.

\section{Elementary bound:}
Consider the quantity
\begin{eqnarray}
X(t) = T(t) \; T(t)^T &=& \mathscr{T} \left\{ \exp\left( 
\int_{t_i}^t 
\left[\begin{array}{cc} 0 & \omega_0\\ -\omega^2(\bar t)/\omega_0 & 0 \end{array}\right] 
\d \bar t
\right) \right\}  
\nonumber\\
&& \times 
\left[\mathscr{T} \left\{ \exp\left( 
\int_{t_i}^t 
\left[\begin{array}{cc} 0 & \omega_0\\ -\omega^2(\bar t)/\omega_0 & 0 \end{array}\right] 
\d \bar t
\right) \right\} \right]^T.
\end{eqnarray}
This object satisfies the differential equation
\begin{equation}
{\d X\over\d t} =  
\left[\begin{array}{cc} 0 & \omega_0\\ -\omega^2(\bar t)/\omega_0 & 0 \end{array}\right]  \; X(t)  
+
X(t)\;  \left[\begin{array}{cc} 0 &-\omega^2(\bar t)/ \omega_0\\ \omega_0 & 0 \end{array}\right],
\end{equation}
with the boundary condition
\begin{equation}
X(t_i)=I.
\end{equation}
Now note
\begin{equation}
\tr(X) = \tr\{ T \, T^T\} = a^2 +b^2 + c^2 + d^2.
\end{equation}
Furthermore 
\begin{eqnarray}
{\d X\over \d t} 
&=& 
 \left[\begin{array}{cc} 0 & \omega_0\\ -\omega^2/\omega_0 & 0 \end{array}\right]  
\left[\begin{array}{cc} a^2 + b^2 & ac + bd\\ ac + bd & c^2 + d^2 \end{array}\right] 
\nonumber
\\
&& \qquad\qquad 
+ \left[\begin{array}{cc} a^2 + b^2 & ac + bd\\ ac + bd & c^2 + d^2 \end{array}\right]  
\left[\begin{array}{cc} 0 &-\omega^2/ \omega_0\\ \omega_0 & 0 \end{array}\right],
\qquad\qquad
\\
&&=\left[\begin{array}{cc} 2 \omega_0(ac + bd)  & 
\omega_0(c^2 + d^2) - (\omega^2/\omega_0) (a^2 + b^2)\\ 
\omega_0(c^2 + d^2) - (\omega^2/\omega_0) (a^2 + b^2) & 
(-2 \omega^2/\omega_0) (ac + bd)\end{array}\right],
\nonumber
\end{eqnarray}
and so we see
\begin{equation}
\tr \left\{ \left[\begin{array}{cc} 0 & \omega_0\\ -\omega^2/\omega_0 & 0 \end{array}\right]  X +
X  \left[\begin{array}{cc} 0 &-\omega^2/ \omega_0\\ \omega_0 & 0 \end{array}\right] \right\}
=
2(ac+bd) \left[ \omega_0-{\omega^2\over\omega_0}\right].
\end{equation}
Therefore
\begin{equation}
{\d \tr[X]\over \d t } 
=2(ac+bd) \left[ \omega_0-{\omega^2\over\omega_0}\right].
\end{equation}
Using this key result, and some very simple analysis, we shall now derive our first elementary bound on the Bogoliubov coefficients.
\begin{itemize}
\item 
For any 2 real numbers, using $(x+y)^2\geq 0$ and $(x-y)^2\geq0$, we have
 \begin{equation}
 x^2 + y^2 \geq 2|xy|.
 \end{equation}
 In particular, for any 4 real numbers  this implies
 \begin{equation}
 a^2+b^2+c^2+d^2 \geq 2 \sqrt{(a^2+b^2)(c^2+d^2)}.
 \end{equation}
 
 \item
 But we also have
 \begin{eqnarray}
|ac+bd|^2 + |ad-bc|^2 &=& a^2 c^2 + 2 abcd +b^2d^2 +a^2d^2 -2 abcd+ b^2 c^2 \quad
\\
& =&  (a^2+b^2)(c^2+d^2), 
\end{eqnarray}
thus, for any 4 real numbers
\begin{equation}
 a^2 +b^2 + c^2 +d^2 \geq 2 \sqrt{|ac+bd|^2+|ad-bc|^2}.
 \end{equation} 

 \item
For the particular case we are interested in we additionally have the unit determinant condition $ad-bc=1$, so the above implies
\begin{equation}
 a^2 +b^2 + c^2 +d^2 \geq 2 \sqrt{|ac+bd|^2+1},
\end{equation} 
 whence
 \begin{equation}
2  |ac+bd| \leq \sqrt{ (a^2 +b^2 + c^2 +d^2)^2 - 4}.
\end{equation} 
\end{itemize}
Then
\begin{equation}
{\d \tr[X]\over \d t } 
=2(ac+bd) \left[ \omega_0-{\omega^2\over\omega_0}\right]
\leq 2|ac+bd| \;\left| \omega_0-{\omega^2\over\omega_0}\right|,
\end{equation}
whence
\begin{equation}
{\d \tr[X]\over \d t } \leq  \sqrt{ (a^2 +b^2 + c^2 +d^2)^2 - 4} \;
\left| \omega_0-{\omega^2\over\omega_0}\right| = 
\sqrt{\tr[X]^2-4}  \; \left| \omega_0-{\omega^2\over\omega_0}\right|,
\end{equation}
whence
\begin{equation}
{1\over \sqrt{\tr[X]^2-4}} \; {\d \tr[X]\over \d t } \leq  \left| \omega_0-{\omega^2\over\omega_0}\right|.
\end{equation}
This implies
\begin{equation}
{\d \cosh^{-1}\tr[X/2]\over\d t} \leq  \left| \omega_0-{\omega^2\over\omega_0}\right|,
\end{equation}
whence
\begin{equation}
\tr[X] \leq 2\cosh\left\{ \int _{t_i}^{t_f} \left| \omega_0-{\omega^2\over\omega_0}\right| \d t \right\}.
\end{equation}
We now have
 \begin{equation}
 |\beta|^2 =  {1\over 4} \left\{  \tr\left\{  T \;  T^T \right\} - 2 \right\} 
 =   {1\over 4} \left\{  \tr\left\{  X \right\} - 2 \right\},
 \end{equation}
so that
 \begin{eqnarray}
 |\beta|^2 
 &\leq&
 {1\over 2} \left\{ \cosh\left\{ \int _{t_i}^{t_f} \left| \omega_0-{\omega^2\over\omega_0}\right| \d t \right\} -1\right\},
\\
&=& 
\sinh^2\left\{ {1\over2} \int _{t_i}^{t_f} \left| \omega_0-{\omega^2\over\omega_0}\right| \d t \right\}.
 \end{eqnarray}
 So finally
\begin{equation}
 |\beta|^2  \leq   
 \sinh^2\left\{ {1\over2} \int _{t_i}^{t_f} \left| \omega_0-{\omega^2\over\omega_0}\right| \d t \right\},
\end{equation}
and consequently
\begin{equation}
 |\alpha|^2  \leq   
 \cosh^2\left\{ {1\over2} \int _{t_i}^{t_f} \left| \omega_0-{\omega^2\over\omega_0}\right| \d t \right\}.
\end{equation}
These bounds are quite remarkable in their generality. A version of this result was derived in~\cite{bounds} but the present derivation is largely independent and has the virtue of being utterly elementary --- in particular, the use of complex numbers has been minimized, and we have completely eliminated  the use of the ``auxiliary functions''  and ``gauge conditions'' that were needed for the derivation in~\cite{bounds} .

If one translates this to the space domain, then the equivalent barrier penetration coefficient is $T_\mathrm{transmission} \leftrightarrow 1/|\alpha|^2$, and the equivalent reflection coefficient is  $R \leftrightarrow |\beta^2|/|\alpha|^2$. Making the appropriate translations
\begin{equation}
T_\mathrm{transmission} \geq 
\sech^2\left\{ {1\over2} \int _{x_i}^{x_f} \left| k_0-{k^2(x)\over k_0}\right| \d x \right\},
\end{equation}
and 
\begin{equation}
R \leq \tanh^2\left\{ {1\over2} \int _{x_i}^{x_f} \left| k_0-{k^2(x)\over k_0}\right| \d x \right\}.
\end{equation}
(For completeness we mention that reference~\cite{bounds} provides a number of consistency checks on these bounds by comparing them with known exact results~\cite{exact}.)

\section{Lower bound on $|\beta|^2$}

To obtain a lower bound on the $|\beta|$ Bogoliubov coefficient, consider any real valued parameter $\epsilon$. Then since the matrix $T$ is itself real,
\begin{equation}
\tr \left\{(T -\epsilon\, T^T)^T\;(T-\epsilon\, T^T)\right\} \geq 0,
\end{equation}
so that
\begin{equation}
(1+\epsilon^2) \, \tr (T \, T^T) - 2  \epsilon \,  \tr(T^2)  \geq 0,
\end{equation}
whence
\begin{equation}
\tr (T^T \, T) \geq {2  \epsilon\over1+\epsilon^2} \; \tr (T^2), 
\end{equation}
This bound is extremized for $\epsilon=\pm1$, whence
\begin{equation}
\tr (T^T \,T) \geq  \left|\tr (T^2)\right|, 
\end{equation}
and so
\begin{equation}
|\beta|^2 \geq {1 \over 4} \left\{  \left|\tr (T^2)\right| - 2 \right\}.
\end{equation}
This is certainly a bound, but it is not as useful as one might hope. It is useful only if $\tr[T^2]>2$. But
\begin{equation}
\tr[T^2] = a^2+d^2+2bc = a^2+d^2 +2(ad-1) = (a+d)^2-2 = (\tr[T])^2-2.
\end{equation}
So using the unit determinant condition,  $\tr[T^2]>2$ can be seen to require $|a+d|\geq2$, that is, $\tr[T]>2$. But when does this happen? For the real matrix
\begin{equation}
\left[\begin{array}{cc} a & b\\ c & d \end{array}\right] 
\end{equation}
with unit determinant the eigenvalues are
\begin{equation}
\lambda = {a+d\over2} \pm {\sqrt{ (a+d)^2 - 4}\over 2}.
\end{equation}
The condition $a+d>2$ is thus equivalent to the condition that the eigenvalues are real.  Unfortunately there seems to be no simple way to then relate this to the properties of the function $\omega(t)$.

\section{A more general upper bound}

Now let $\Omega(t)$ be an \emph{arbitrary} everywhere real and nonzero \emph{function} of $t$ with the dimensions of frequency. Then we can rewrite the Schrodinger ODE (\ref{E:time}) as:
\begin{equation}
{\d\over\d t} \left[\begin{array}{c} \phi \; \sqrt{\Omega}\\ \pi/\sqrt{\Omega} \end{array}\right] =
\left[\begin{array}{cc} 
{1\over2}(\dot\Omega/\Omega) & \Omega\\
 -\omega^2(t)/\Omega & -{1\over2}(\dot\Omega/\Omega) 
 \end{array}\right] \; 
\left[\begin{array}{c} \phi\;\sqrt{\Omega}\\ \pi/\sqrt{\Omega} \end{array}\right].
\end{equation}
Again all the matrix elements have been carefully chosen to have the same engineering dimension. We can formally solve this in terms of the time-ordered product:
\begin{equation}
\left[\begin{array}{c} \phi \, \sqrt{\Omega}\\ \pi/\sqrt{\Omega} \end{array}\right]_t = 
\mathscr{T} \left\{ \exp\left( 
\int_{t_0}^t \left[\begin{array}{cc} 
{1\over2}(\dot\Omega/\Omega)  & \Omega\\ 
-\omega^2(\bar t)/\Omega & - {1\over2}(\dot\Omega/\Omega) 
\end{array}\right] \d \bar t
\right) \right\} \; \left[\begin{array}{c} \phi\\ \pi/\sqrt{\Omega} \end{array}\right]_{t_0}.
\end{equation}
The new $T$ matrix is
\begin{equation}
T = \mathscr{T} \left\{ \exp\left( 
\int_{t_i}^{t_f}\left[\begin{array}{cc} 
{1\over2}(\dot\Omega/\Omega)  & \Omega\\ 
-\omega^2(\bar t)/\Omega & -{1\over2}(\dot\Omega/\Omega) 
 \end{array}\right] \d \bar t
\right) \right\}.
\end{equation}
Note that the matrix $T$ is still real, and that because
\begin{equation}
\tr \left[\begin{array}{cc} 
{1\over2}(\dot\Omega/\Omega)  & \Omega\\ 
-\omega^2(\bar t)/\Omega & -{1\over2}(\dot\Omega/\Omega) 
 \end{array}\right] 
 =0
 \end{equation}
 it still follows that $T$ has determinant unity:
 \begin{equation}
 T = \left[\begin{array}{cc} 
a & b\\ 
c & d
 \end{array}\right]; \qquad   ad-bc = 1.
\end{equation}
This means that much of the earlier computations carry through without change. In particular as long as at the initial and final times we impose $\Omega(t)\to\omega_0$ as $t\to t_f$ and $t\to t_i$, we still have
\begin{equation}
\alpha = {1\over2} \left[a+ d 
+ i\left({ b} -{ c}\right)\right] 
\exp(-i \omega_0[t_f - t_i]),
\end{equation}
\begin{equation}
\beta = {1\over2} \left[ a- d 
+ i\left({b} +{c}\right)\right] 
\exp(-i \omega_0[t_f + t_i]),
\end{equation}
\begin{equation}
|\beta|^2 =  {1\over 4} \; \tr\left\{  T \;  T^T  - I\right\},
\end{equation} 
\begin{equation}
|\alpha|^2 =  {1\over 4} \; \tr\left\{   T \;  T^T  + I\right\}.
\end{equation} 
Now consider the quantity
\begin{eqnarray}
X(t) = T(t) \; T(t)^T &=& \mathscr{T} \left\{ \exp\left( 
\int_{t_i}^t \left[\begin{array}{cc} 
{1\over2}(\dot\Omega/\Omega)  & \Omega\\ 
-\omega^2(\bar t)/\Omega & -{1\over2}(\dot\Omega/\Omega) 
 \end{array}\right] \d \bar t
\right) \right\}  
\nonumber
\\
&&\times
\left[\mathscr{T} \left\{ \exp\left( 
\int_{t_i}^t \left[\begin{array}{cc} 
{1\over2}(\dot\Omega/\Omega)  & \Omega\\ 
-\omega^2(\bar t)/\Omega & -{1\over2}(\dot\Omega/\Omega) 
 \end{array}\right] \d \bar t
\right) \right\}  \right]^T\!\!.\;\;
\end{eqnarray}
This now satisfies the differential equation
\begin{equation}
{\d X\over\d t} =  
\left[\begin{array}{cc} 
{1\over2}(\dot\Omega/\Omega)  & \Omega\\ 
-\omega^2(\bar t)/\Omega & -{1\over2}(\dot\Omega/\Omega)  
\end{array}\right]  X +
X  \left[\begin{array}{cc} 
{1\over2}(\dot\Omega/\Omega)  &-\omega^2(\bar t)/ \Omega\\ 
\Omega & -{1\over2}(\dot\Omega/\Omega)  \end{array}\right],
\end{equation}
with the boundary condition
\begin{equation}
X(t_i)=I,
\end{equation}
and
\begin{equation}
\tr[X] = a^2+b^2+c^2+d^2.
\end{equation}
A brief computation yields
\begin{eqnarray}
{\d X\over\d t} &=&  
\left[\begin{array}{cc} 
{1\over2}(\dot\Omega/\Omega)  & \Omega\\ 
-\omega^2(\bar t)/\Omega & -{1\over2}(\dot\Omega/\Omega)  
\end{array}\right]  \left[\begin{array}{cc} a^2 + b^2 & ac + bd\\ ac + bd & c^2 + d^2 \end{array}\right] 
\nonumber\\
&&
 +  \left[\begin{array}{cc} a^2 + b^2 & ac + bd\\ ac + bd & c^2 + d^2 \end{array}\right] 
 \left[\begin{array}{cc} 
{1\over2}(\dot\Omega/\Omega)  &-\omega^2(\bar t)/ \Omega\\ 
\Omega & -{1\over2}(\dot\Omega/\Omega)  \end{array}\right],
\end{eqnarray}
\begin{equation}
= \left[\begin{array}{cc} 
(\dot\Omega/\Omega)(a^2 + b^2) + 2 \Omega(ac + bd)  & \Omega (c^2 + d^2) - (\omega^2 / \Omega) (a^2 + b^2)\\ 
- (\omega^2/\Omega) (a^2 + b^2) + \Omega (c^2 + d^2) & - (2 \omega^2 / \Omega) (ac + bd) - (\dot \Omega/\Omega)(c^2 + d^2)
\end{array}\right].
\end{equation}
Then taking the trace, there is now one extra term
\begin{equation}
{\d\tr[X]\over\d t} =   
(a^2+b^2-c^2-d^2)\left[{\dot\Omega\over\Omega}\right] +   
2(ac+bd) \left[ \Omega-{\omega^2\over\Omega}\right]
\end{equation} 
Note that if $\Omega(t)\to\omega_0$ then $\dot\Omega\to 0$ and we recover the ODE of the ``elementary'' bound.  In this more general setting we now proceed by using the following facts: 
\begin{itemize}
\item 
As previously we note
 \begin{equation}
|ac+bd|^2 + |ad-bc|^2 = a^2 c^2 + 2 abcd +b^2d^2 +a^2d^2 -2 abcd+ b^2 c^2 =  (a^2+b^2)(c^2+d^2),
\end{equation}
which implies
\begin{equation}
|ac+bd| = \sqrt{ (a^2+b^2)(c^2+d^2) -1},
\end{equation}
that is
\begin{equation}
2 |ac+bd| = \sqrt{ 4 (a^2+b^2)(c^2+d^2) -4}.
\end{equation}
\item
Additionally, we use
\begin{equation}
|a^2+b^2-c^2-d^2| = \sqrt{|a^2+b^2+c^2+d^2|^2- 4 (a^2+b^2)(c^2+d^2)},
\end{equation}
implying
\begin{equation}
|a^2+b^2-c^2-d^2| ^2 + (2 |ac+bd| )^2 = |a^2+b^2+c^2+d^2|^2- 4.
\end{equation}
\end{itemize}
In particular, combining these observations,  this means that we can find an angle $\theta$ (which is in general some complicated real function of $a$, $b$, $c$, $d$) such that
\begin{equation}
2 (ac+bd) = \sqrt{ |a^2+b^2+c^2+d^2|^2- 4} \; \;\sin\theta,
\end{equation}
\begin{equation}
a^2+b^2-c^2-d^2 = \sqrt{ |a^2+b^2+c^2+d^2|^2- 4} \; \; \cos\theta,
\end{equation}
whence
\begin{equation}
{\d\tr[X]\over\d t} =   \sqrt{ |a^2+b^2+c^2+d^2|^2- 4} \; 
\left\{ \sin\theta\left[{\dot\Omega\over\Omega}\right] +   
\cos\theta \left[ \Omega-{\omega^2\over\Omega}\right] \right\}.
\end{equation} 
But for any real $\theta$ we certainly have the \emph{inequality}
\begin{equation}
\sin\theta\left[{\dot\Omega\over\Omega}\right] +   
\cos\theta \left[ \Omega-{\omega^2\over\Omega}\right] 
\leq 
\sqrt{ \left[{\dot\Omega\over\Omega}\right]^2 + \left[ \Omega-{\omega^2\over\Omega}\right]^2},
\end{equation}
implying
\begin{equation}
{\d\tr[X]\over\d t} \leq   \sqrt{ |a^2+b^2+c^2+d^2|^2- 4} \; \;
\sqrt{ \left[{\dot\Omega\over\Omega}\right]^2 + \left[ \Omega-{\omega^2\over\Omega}\right]^2}.
\end{equation}
Therefore
\begin{equation}
{\d\tr[X]\over\d t} \leq   \sqrt{ \tr[X]^2- 4} \; 
\sqrt{ \left[{\dot\Omega\over\Omega}\right]^2 + \left[ \Omega-{\omega^2\over\Omega}\right]^2}
\end{equation}
implying
\begin{equation}
{1\over \sqrt{ \tr[X]^2- 4} } {\d\tr[X]\over\d t}  \leq
\sqrt{ \left[{\dot\Omega\over\Omega}\right]^2 + \left[ \Omega-{\omega^2\over\Omega}\right]^2},
\end{equation}
whence
\begin{equation}
 {\d\cosh^{-1}(\tr[X]/2)\over\d t}  \leq
\sqrt{ \left[{\dot\Omega\over\Omega}\right]^2 + \left[ \Omega-{\omega^2\over\Omega}\right]^2},
\end{equation}
so that 
\begin{equation}
\tr[X] = \tr[T \;T^T] \leq 2 \cosh\left\{ \lint_{\!\!\!\!\!\!t_i}^{t_f} 
\sqrt{ \left[{\dot\Omega\over\Omega}\right]^2 + \left[ \Omega-{\omega^2\over\Omega}\right]^2}
\; \d t \right\}.
\end{equation}
Using the general formulae for $|\alpha|^2$ and $|\beta^2|$ in terms of $\tr\{T \, T^T\}$, and simplifying, we see
\begin{equation}
|\beta|^2 \leq  \sinh^2\left\{ {1\over 2} \int_{t_i}^{t_f} 
{1 \over |\Omega|}\sqrt{ \dot\Omega^2 + \left[ \Omega^2-{\omega^2}\right]^2}
\; \d t \right\},
\end{equation}
and
\begin{equation}
|\alpha|^2 \leq  \cosh^2\left\{ {1\over 2} \int_{t_i}^{t_f} 
{1 \over |\Omega|}\sqrt{ \dot\Omega^2 + \left[ \Omega^2-{\omega^2}\right]^2}
\; \d t \right\}.
\end{equation}
This result is completely equivalent to the corresponding result in~\cite{bounds}; though again note that the derivation is largely independent and that it no longer requires one to introduce any ``gauge fixing'' condition, nor need we introduce any WKB-like ansatz. The current proof is much more ``direct", and at worst uses simple inequalities and  straightforward ODE theory.
If we work in the space domain instead of the time domain and make the translations $\Omega(t)\to \varphi'(x)$, $\omega(t) \to k(x)$, we see
\begin{equation}
|\alpha|^2 \leq  \cosh^2\left\{ {1\over 2} \int_{x_i}^{x_f} 
{1 \over |\varphi'|}\sqrt{ (\varphi'')^2 + \left[ (\varphi')^2-{k^2}\right]^2}
\; \d x \right\},
\end{equation}
and
\begin{equation}
|\beta|^2 \leq  \sinh^2\left\{ {1\over 2} \int_{x_i}^{x_f} 
{1 \over |\varphi'|}\sqrt{ (\varphi'')^2 + \left[ (\varphi')^2-{k^2}\right]^2}
\; \d x \right\}.
\end{equation}
This is perhaps physically more transparent in terms of the equivalent transmission and reflection coefficients
\begin{equation}
T_\mathrm{transmission} \geq  \sech^2\left\{ {1\over 2} \int_{x_i}^{x_f} 
{1 \over |\varphi'|}\sqrt{ (\varphi'')^2 + \left[ (\varphi')^2-{k^2}\right]^2}
\; \d x \right\},
\end{equation}
and
\begin{equation}
R \leq  \tanh^2\left\{ {1\over 2} \int_{x_i}^{x_f} 
{1 \over |\varphi'|}\sqrt{ (\varphi'')^2 + \left[ (\varphi')^2-{k^2}\right]^2}
\; \d x \right\}.
\end{equation}
(For completeness we mention that reference~\cite{bounds} provides a number of consistency checks on these bounds by comparing them with known exact results~\cite{exact}.)

\section{The ``optimal'' choice of $\Omega(t)$?}

What is the \emph{optimal} choice of $\Omega(t)$ that one can make? Leading to the most stringent bound on the Bogoliubov coefficients? The bound we have just derived holds for arbitrary $\Omega(t)$, subject to the two boundary conditions $\Omega(t_i) = \omega_0 = \Omega(t_f)$ and the overall constraint $\Omega(t) \neq 0$. Since $\sinh$ and $\cosh$ are both convex functions, finding the most stringent constraint on $|\beta|$ and $|\alpha|$ is thus a variational calculus problem equivalent to minimizing the action
\begin{equation}
S =  \int_{t_i}^{t_f} 
{1 \over |\Omega|}\;\sqrt{ \dot\Omega^2 + \left[ \Omega^2-{\omega^2}\right]^2}
\; \d t.
\end{equation}
The relevant Euler--Lagrange equations are quite messy, and progress (at least insofar as there is any practicable progress) is better made by using an indirect attack. The Lagrangian is 
\begin{equation}
L =  
{1 \over |\Omega|}\; \sqrt{ \dot\Omega^2 + \left[ \Omega^2-{\omega^2}\right]^2},
\end{equation}
and so the corresponding canonical momentum can be evaluated as
\begin{equation}
\pi = {\partial L\over\partial \dot\Omega} =
 {\dot \Omega \over |\Omega|\;\sqrt{ \dot\Omega^2 + \left[ \Omega^2-{\omega^2}\right]^2}}.
\end{equation}
From the boundary conditions we can deduce
\begin{equation}
\pi(t_i) = {1\over\omega_0} = \pi(t_f).
\end{equation}
The Hamiltonian is now
\begin{equation}
H = \pi\; \dot \Omega - L = 
 {\dot \Omega^2 - \left\{ \dot\Omega^2 + \left[ \Omega^2-{\omega^2}\right]^2 \right\}
 \over |\Omega|\;\sqrt{ \dot\Omega^2 + \left[ \Omega^2-{\omega^2}\right]^2}}
 =
-  {\left[ \Omega^2-{\omega^2}\right]^2
 \over |\Omega|\;\sqrt{ \dot\Omega^2 + \left[ \Omega^2-{\omega^2}\right]^2}}.
\end{equation}
Unfortunately the Hamiltonian is \emph{explicitly} time-dependent [via $\omega(t)$] and so is \emph{not} conserved. The best we can say is that at the endpoints of the motion
\begin{equation}
H(t_i) = 0 = H(t_f).
\end{equation}
By solving for $\dot\Omega$ as a function of $\pi$ and $\Omega$ we can also write
\begin{equation}
\dot \Omega = {\pi \,\Omega \;\over \sqrt{1- \pi^2 \,\Omega^2}} \;  (\Omega^2-\omega^2),
\end{equation}
and
\begin{equation}
H = - {\sqrt{1-\pi^2\,\Omega^2} \;(\Omega^2-\omega^2)\over |\Omega|}.
\end{equation}
Note that $\dot\Omega$ at the endpoints is \emph{cannot} in general be explicitly evaluated in terms of the boundary conditions. 

An alternative formulation which slightly simplifies the analysis is to change variables by writing
\begin{equation}
\Omega(t) = \omega_0 \; \exp[\theta(t)],
\end{equation}
where the boundary conditions are now
\begin{equation}
\theta(t_i) = 0 = \theta(t_f),
\end{equation}
and the action is now rewritten as
\begin{equation}
S =  \int_{t_i}^{t_f} 
\sqrt{ \dot\theta^2 + \omega_0^2 \left[ e^{2\theta} -{\omega^2\over\omega_0^2}\; e^{-2\theta} \right]^2}
\; \d t.
\end{equation}
Then, in terms of this new variable we have
\begin{equation}
L =  
\sqrt{ \dot\theta^2 + \omega_0^2 \left[ e^{2\theta} -{\omega^2\over\omega_0^2}\; e^{-2\theta} \right]^2},
\end{equation}
with (dimensionless) conjugate momentum 
\begin{equation}
\pi = {\partial L\over\partial \dot\theta} =
 {\dot \theta \over \sqrt{ \dot\theta^2 
 +  \omega_0^2 \left[ e^{2\theta} -{\omega^2\over\omega_0^2}\; e^{-2\theta} \right]^2}},
\end{equation}
and boundary conditions
\begin{equation}
\pi(t_i) = 1 = \pi(t_f).
\end{equation}
The (non-conserved) Hamiltonian is
\begin{equation}
H = \pi\; \dot \theta - L = 
-  {\omega_0^2 \left[ e^{2\theta} -{\omega^2\over\omega_0^2}\; e^{-2\theta} \right]^2
 \over \sqrt{ \dot\theta^2 + \omega_0^2 \left[ e^{2\theta} -{\omega^2\over\omega_0^2}\; e^{-2\theta} \right]^2}},
\end{equation}
subject to
\begin{equation}
H(t_i) = 0 = H(t_f).
\end{equation}
Inverting, we see
\begin{equation}
\dot \theta = {\pi  \over \sqrt{1- \pi^2}} \; \omega_0 \;
\left[ e^{2\theta} -{\omega^2\over\omega_0^2}\; e^{-2\theta} \right] ,
\end{equation}
and
\begin{equation}
H = - \sqrt{1-\pi^2} \;  \omega_0 \;
\left[ e^{2\theta} -{\omega^2\over\omega_0^2}\; e^{-2\theta} \right].
\end{equation}
This has given us a somewhat simpler variational problem,  unfortunately the Euler--Lagrange equations are still too messy to provide useful results.

Overall, we see that while solving the variational problem would indeed result in an optimum bound, there is no explicit general formula for such a solution. In the tradeoff between optimality and explicitness, we will have to accept the use of sub-optimal but explicit bounds.

\section{Sub-optimal but explicit bounds}

From our general bounds
\begin{equation}
|\beta|^2 \leq  \sinh^2\left\{ {1\over 2} \int_{t_i}^{t_f} 
{1 \over |\Omega|}\sqrt{ \dot\Omega^2 + \left[ \Omega^2-{\omega^2}\right]^2}
\; \d t \right\},
\end{equation}
and
\begin{equation}
|\alpha|^2 \leq  \cosh^2\left\{ {1\over 2} \int_{t_i}^{t_f} 
{1 \over |\Omega|}\sqrt{ \dot\Omega^2 + \left[ \Omega^2-{\omega^2}\right]^2}
\; \d t \right\},
\end{equation}
the following special cases are of particular interest:
\begin{description}
\item[$\Omega=\omega_0$:]  In this case we simply obtain the ``elementary'' bound considered above.

\item[$\Omega=\omega$:]  This case only makes sense if $\omega^2>0$ is always positive. (Otherwise $\omega$ and hence $\Omega$ becomes imaginary in the ``classically forbidden'' region; the matrix $T$ then becomes complex, and the entire formalism breaks down). Subject to this constraint we find
\begin{equation}
|\beta|^2 \leq  \sinh^2\left\{ {1\over 2} \int_{t_i}^{t_f} 
\left|{\dot \omega \over \omega} \right|
\; \d t \right\},
\end{equation}
and
\begin{equation}
|\alpha|^2 \leq  \cosh^2\left\{ {1\over 2} \int_{t_i}^{t_f} 
\left|{\dot \omega \over \omega} \right|
\; \d t \right\}.
\end{equation}
This case was also considered in~\cite{bounds}.

\item[$\Omega=\omega^\epsilon \;\omega_0^{1-\epsilon}$:]  This case again only makes sense if $\omega^2>0$ is always positive. Subject to this constraint we find
\begin{equation}
|\beta|^2 \leq  \sinh^2\left\{ {1\over 2} \lint_{\!\!\!\!\!\!t_i}^{t_f} 
\sqrt{ \epsilon^2 \; {\dot\omega^2\over\omega^2}  + {
\omega^{2\epsilon} \left[ \omega_0^{2-2\epsilon}-{\omega^{2-2\epsilon}}\right]^2\over \omega_0^{2-2\epsilon}}}
\; \d t \right\},
\end{equation}
and
\begin{equation}
|\alpha|^2 \leq  \cosh^2\left\{ {1\over 2} \lint_{\!\!\!\!\!\!t_i}^{t_f} 
\sqrt{ \epsilon^2 \; {\dot\omega^2\over\omega^2}  + {
\omega^{2\epsilon} \left[ \omega_0^{2-2\epsilon}-{\omega^{2-2\epsilon}}\right]^2\over \omega_0^{2-2\epsilon}}}
\; \d t \right\}.
\end{equation}
This nicely interpolates between the two cases given above, which correspond to $\epsilon=0$ and $\epsilon=1$ respectively. 

\item[Triangle inequality:] Since $\sqrt{x^2+y^2} \leq |x| + |y|$ we see that
\begin{equation}
|\beta|^2 \leq  \sinh^2\left\{ {1\over 2} \int_{t_i}^{t_f} 
\left|{\dot \Omega \over \Omega} \right| \; \d t
+ 
 {1\over 2} \int_{t_i}^{t_f} 
\left| \Omega-{\omega^2\over\Omega}\right|
\; \d t \right\},
\end{equation}
and
\begin{equation}
|\alpha|^2 \leq  \cosh^2\left\{ {1\over 2} \int_{t_i}^{t_f} 
\left|{\dot \Omega \over \Omega} \right| \; \d t
+ 
 {1\over 2} \int_{t_i}^{t_f} 
\left| \Omega-{\omega^2\over\Omega}\right|
\; \d t \right\}.
\end{equation}
\end{description}
These bounds, because they are explicit, are often the most useful quantities to calculate.

\section{The ``interaction picture''}

If we split the function $\omega(t)^2$ into an exactly solvable piece $\omega_e(t)^2$ and a perturbation $\omega_\Delta(t)^2$ then we can develop a formal perturbation series for the transfer matrix $T$, in close analogy to the procedures for developing quantum field theoretic  perturbation theory in the interaction picture. Specifically let us write
\begin{equation}
\omega_e(t)^2 = \omega_e(t)^2 + \omega_\Delta(t)^2,
\end{equation}
and
\begin{equation}
{\d T(t)\over\d t} = Q(t) \; T(t) = \left[Q_e(t) + Q_\Delta(t) \right] \; T(t).
\end{equation}
Now defining
\begin{equation}
T(t) = T_e(t) \; T_\Delta(t),
\end{equation}
we shall develop a formal solution for $T_\Delta(t)$. Consider
\begin{equation}
{\d T(t)\over\d t} =  \left[Q_e(t) + Q_\Delta(t) \right] \; T_e(t) \; T_\Delta(t),
\end{equation}
and compare it with
\begin{equation}
{\d T(t)\over\d t} = {\d T_e (t)\over\d t}\; T_\Delta(t)  + T_e(t) \; {\d T_\Delta(t)\over\d t} = 
 Q_e(t) \; T_e(t) \; T_\Delta(t)  +  T_e(t) \; {\d T_\Delta(t)\over\d t}.
 \end{equation}
 Therefore
 \begin{equation}
 {\d T_\Delta(t)\over\d t} = \left\{ T_e(t)^{-1} \;  Q_\Delta(t) \; T_e(t) \right\} \; T_\Delta,
\end{equation}
whence
\begin{equation}
T_\Delta(t) = \mathscr{T} \exp\left( \int_{t_i}^t \left\{ T_e(\bar t)^{-1} \;  Q_\Delta(\bar t) \; T_e(\bar t) \right\}  \d \bar t \right).
\end{equation}
For the full transfer matrix $T$ we have
\begin{equation}
T(t) = T_e(t) \times \mathscr{T} \exp\left( \int_{t_i}^t \left\{ T_e(\bar t)^{-1} \;  Q_\Delta(\bar t) \; T_e(\bar t) \right\}  \d \bar t \right),
\end{equation}
and we have succeeded into splitting it into an exact piece $T_e(t)$ plus a distortion due to $Q_\Delta(t)$. This can now be used as the starting point for a perturbation expansion. (The analogy with quantum field theoretic perturbation theory in the interaction picture should now be completely clear.)

To develop some formal bounds on the Bogoliubov coefficients it is useful to suppress (currently) unnecessary phases by defining
\begin{equation}
\tilde \alpha = {1\over2} \left[a+ d 
+ i\left({ b} -{ c}\right)\right] ,
\end{equation}
\begin{equation}
\tilde \beta = {1\over2} \left[ a- d 
+ i\left({b} +{c}\right)\right].
\end{equation}
The virtue of these definitions is that for $T = T_e\; T_\Delta$ they satisfy a simple composition rule which can easily  be verified via matrix multiplication. From $T = T_e\; T_\Delta$ we have
\begin{equation}
\left[
\begin{array}{cc}
a  & b   \\
c  &  d 
\end{array}
\right]
=
\left[
\begin{array}{ccc}
a_e \,a_\Delta + b_e \,c_\Delta  &    a_e \,b_\Delta + b_e \,d_\Delta \\
c_e \,a_\Delta + d_e \,c_\Delta   &    c_e \,b_\Delta + d_e \,d_\Delta
\end{array}
\right].
\end{equation}
Then some simple linear algebra leads to
\begin{equation}
\tilde\beta = \tilde \alpha_e \; \tilde \beta_\Delta + \tilde \beta_e \; \tilde \alpha^*_\Delta,
\end{equation}
\begin{equation}
\tilde\alpha = \tilde \alpha_e \; \tilde \alpha_\Delta + \tilde \beta_e \; \tilde \beta^*_\Delta,
\end{equation}
But then
\begin{equation}
|\beta| = |\tilde\beta| =  \left|\tilde \alpha_e \; \tilde \beta_\Delta + \tilde \beta_e \; \tilde \alpha^*_\Delta\right|
\leq  \left|\tilde \alpha_e \; \tilde \beta_\Delta\right| + \left| \tilde \beta_e \; \tilde \alpha^*_\Delta\right|
=  \left| \alpha_e \; \beta_\Delta\right| + \left| \beta_e \;  \alpha_\Delta\right|,
\end{equation}
that is
\begin{equation}
|\beta| \leq  \left| \alpha_e \right|\; \left|\beta_\Delta\right| + \left| \beta_e \right|\;  \left|\alpha_\Delta\right|,
\end{equation}
or the equivalent
\begin{equation}
|\beta| \leq  \sqrt{1+\left| \beta_e \right|^2} \; \left|\beta_\Delta\right| + \left| \beta_e \right|\;  \sqrt{1+ \left|\beta_\Delta\right|^2}.
\end{equation}
Similarly
\begin{equation}
|\beta| = |\tilde\beta| =  \left|\tilde \alpha_e \; \tilde \beta_\Delta + \tilde \beta_e \; \tilde \alpha^*_\Delta\right|
\geq  \left| \; \left|\tilde \alpha_e \; \tilde \beta_\Delta\right| - \left| \tilde \beta_e \; \tilde \alpha^*_\Delta\right|\; \right|
=  \left| \; \left| \alpha_e \; \beta_\Delta\right| - \left| \beta_e \;  \alpha_\Delta\right| \; \right|,
\end{equation}
that is
\begin{equation}
|\beta| \geq  \left| \; \left| \alpha_e \right|\; \left|\beta_\Delta\right| - \left| \beta_e \right|\;  \left|\alpha_\Delta\right| \; \right|,
\end{equation}
or the equivalent
\begin{equation}
|\beta| \geq  \left| \; \sqrt{1+\left| \beta_e \right|^2} \; \left|\beta_\Delta\right| - \left| \beta_e \right|\;  \sqrt{1+ \left|\beta_\Delta\right|^2} \; \right|.
\end{equation}
The benefit now is that one has bounded the Bogoliubov coefficient in terms of the (assumed known) exact coefficient $\beta_e$ and the contribution from the perturbation $\beta_\Delta$. Suitably choosing the split between exact and perturbative contributions to $\omega^2$, one could in principle obtain arbitrarily accurate bounds.

\section{Discussion}

In this article we have re-assessed the general bounds on the Bogoliubov coefficients developed in~\cite{bounds}, providing a new and largely independent derivation of the key results that short-circuits much of the technical discussion in~\cite{bounds}. In particular in the current article we do not need to ``gauge fix'', nor do we need to appeal to any WKB-like ansatz to get the discussion started. Furthermore we have seen how to extend the bounds in~\cite{bounds} in several different ways.

Considering the fundamental importance of the questions we are asking, it is remarkable how little work on this topic can currently be found in the literature.  We do not feel that the current bounds are the best that can be achieved, and strongly suspect that it may be possible to develop yet further extensions both to the current formalism, and to the related formalism originally presented in~\cite{bounds}.

Possible extensions might include somehow relaxing the reality constraint on $\Omega(t)$ without damaging too much of the current formalism, a better understanding of the variational problem defining the ``optimal'' bound (thus hopefully leading to an explicit form  thereof), or using several ``probe functions'' [instead of the single function $\Omega(t)$] to more closely bound the Bogoliubov coefficients.

\appendix
\section*{Appendix: Time ordering}
Time-ordered exponentials are a very convenient trick for formally solving certain matrix differential equations. Suppose we have a differential equation of the form
\begin{equation}
{\d U(t) \over \d t} = H(t) \, U(t),
\end{equation}
where $U(t)$ and $H(t)$ are matrices [or more generally linear operators on some vector space] and the matrix $H(t)$ is generally \emph{not} a constant. [So in particular $H(t_1)$ need not commute with $H(t_2)$.] In many settings $H(t)$ will be an anti-Hermitian matrix in which case $U(t)$ would be unitary --- this is not the situation in the current article where the matrix $H(t)$ is real and traceless but non-symmetric.

If $H(t) = H_0$ is a constant then we have the simple solution
\begin{equation}
U(t) = \exp [H_0 \,t] \; U(0).
\end{equation}
If $H(t)$ is a constant then we define the formal process of ``time ordering''
in terms of the exact solution $U(t)$ which we know exists because of standard existence and uniqueness theorems. That is
\begin{equation}
U(t) = \mathscr{T} \left\{ \exp \bigg[\int_0^t H(t') \, \d t'\bigg] \right\} \, U(0),
\end{equation}
which is equivalent to
\begin{equation}
\mathscr{T} \left\{ \exp \bigg[\int_0^t H(t') \, \d t'\bigg] \right\}= U(t) \; U^{-1}(0).
\end{equation}
If we take this as our fundamental definition of time ordering then
\begin{equation}
{\d \over \d t} \, \mathscr{T} \left\{ \exp \bigg[\int_0^t H(t') \, \d t'\bigg]  \right\}
= H(t) \, U(t) \, U^{-1}(0) = H(t) \; \mathscr{T} \left\{ \exp \bigg[\int_0^t H(t') \, \d t'\bigg] \right\}.
\end{equation}
But  by basic notions of Taylor series expansion
\begin{eqnarray}
\mathscr{T} \left\{ \exp \bigg[\int_0^{t + \Delta t} H(t') \, \d t'\bigg] \right\} &=& \{I + H(t) \, \Delta t + O[(\Delta t)^2]\} \, \mathscr{T} \left\{ \exp \bigg[\int_0^t H(t') \, \d t'\bigg] \right\}
\nonumber\\
&=& \exp \, [H(t) \, \Delta t] \, \mathscr{T} \left\{ \exp \bigg[\int_0^t H(t') \, \d t'\bigg]\right\} + O[(\Delta t)^2] .
\nonumber\\
&&
\end{eqnarray}
Let us now bootstrap this result into a general limit formula for the time ordered exponential integral. For simplicity, split the interval $(0, t)$ into $n$ equal segments and evaluate $H(t)$ at the points
\begin{equation}
t_j = t \; {j \over n} \, ; \qquad j \in [0, n - 1],
\end{equation}
then
\begin{eqnarray}
\mathscr{T} \left\{ \exp \bigg[\int_0^t H(t') \, \d t'\bigg] \right\}&=& 
\exp \, [H (t_{n -1}) \, \Delta t] \, \exp \, [H(t_{n - 2}) \ \Delta t] \, . \, . \, .
\\
\nonumber
&& . \, . \, . \exp \, [H(t_1) \, \Delta t] \, \exp \, [H (t_0) \, \Delta t] + O \, \bigg[{1 \over n}\bigg].
\end{eqnarray}
Alternatively
\begin{eqnarray}
\mathscr{T} \left\{ \exp \bigg[\int_0^t H(t') \, \d t'\bigg] \right\}
&=& \lim_{n\to\infty} \exp \, [H(t_{n-1}) \, \Delta t] \, \exp \, [H(t_{n-2}) \, \Delta t] \, . \, . \, .
\\
\nonumber
&& . \, . \, . \exp \, [H(t_1) \, \Delta t] \, \exp \, [H (t_0) \, \Delta t] \, .
\end{eqnarray}
This limiting process should remind you of the way the Riemann integral is defined, except of course that the $H(t_i)$ need not commute with each other so that the order in which the matrix exponentials are multiplied together is critically important. This is why the product is called ``time ordered''. The parameter $t$ can be any real parameter --- in differential geometry it tends to be a parameter along a curve, sometimes an affine parameter, sometimes even arc length, and the product is then sometimes referred to as ``path ordered'', but in general any old parameter would do.

Note what happens if for some reason the $H(t_i)$ do happen to commute with each other. Then for instance
\begin{equation}
\exp \, [H(t_1) \, \Delta t] \, \exp \, [H(t_0) \, \Delta t] \to \exp \, [\{H(t_1) + H(t_0)\} \, \Delta t]
\end{equation}
a result which is \emph{not} true unless the matrices commute. Continuing in this vein, when the matrices do commute we have
\begin{equation}
\mathscr{T} \left\{ \exp \, \bigg[\int_0^t H(t') \, \d t'\bigg] \right\}
\to 
\lim_{n\to\infty} \, \exp \, [\{H(t_{n -1}) + H(t_{n-2}) \, . \, . \, . H(t_1) + H(t_0)\} \, \Delta t] \, .
\end{equation}
But now the argument of the exponential on the RHS really is the usual Riemann integral, so we have
\begin{equation}
\mathscr{T} \left\{ \exp \, \bigg[\int_{0}^{t} H(t') \, \d t'\bigg] \right\}
\to \exp \, \bigg[ \int_{0}^{t} H(t') \, \d t'\bigg] \, .
\end{equation}
That is, the time-ordered integral reduces to the ordinary integral whenever the matrices $H(t)$ commute with each other. (You could also derive this directly from the original differential equation for $U(t)$.)

In some specific quantum mechanical settings you are more likely to consider the slightly different differential equation
\begin{equation}
{\d U(t) \over \d t} = - i H(t) \; U(t),
\end{equation}
where $H(t)$ is now the Hamiltonian operator on an appropriate Hilbert space and $U$ is the unitary time evolution operator. Then
\begin{equation}
U(t) = \mathscr{T} \left\{ \exp \, \bigg[- i \int_0^t H(t') \, \d t'\bigg] \right\} \, U(0),
\end{equation}
but note that there is nothing fundamentally new or different here.


\end{document}